\documentclass[9pt,conference]{IEEEtran}
\IEEEoverridecommandlockouts

\usepackage[bottom=2.1cm, top=1.20cm,left=1.72cm,right=1.72cm]{geometry}

\usepackage{fancyhdr}
\usepackage[normalem]{ulem}
\usepackage{url}
\usepackage{array}
\usepackage{tikz}
\usepackage{makecell}
\usepackage{booktabs}

\usepackage{caption}
\usepackage{amsthm}
\usepackage{amsmath,amsfonts}
\usepackage{algorithmic}
\usepackage{graphicx}
\usepackage{textcomp}
\usepackage[table,xcdraw]{}
\usepackage{booktabs} 
\usepackage{stfloats}
\usepackage[normalem]{ulem}
\usepackage{subfigure}
\usepackage{multirow}
\usepackage{paralist}
\usepackage{balance}
\usepackage{bm}
\usepackage[ruled,linesnumbered]{algorithm2e}
\usepackage{enumitem}
\usepackage{subcaption}
\usepackage{flushend}

\usepackage{url}
\usepackage[hidelinks]{hyperref}

\usepackage[ruled,linesnumbered]{algorithm2e}
\usepackage{cleveref}
\usepackage{colortbl}

\newcommand\name{MEEK}

\graphicspath{{./graphics/}}

\newcommand{\parlabel}[1]{{\noindent\bf #1}}
\newcommand{\eg}{e.g.\xspace}
\newcommand{\ie}{i.e.\xspace}

\newrobustcmd*\circled[1]{\tikz[baseline=(char.base)]{
            \node[shape=circle,draw,inner sep=0.65pt,fill,text=white,minimum size=0.95em] (char) {\textsf{\small #1}};}}

\crefname{figure}{figure}{figures}

\begin{document}
\title{\name: Re-thinking Heterogeneous Parallel Error Detection Architecture for Real-World OoO Superscalar Processors}

\author{\IEEEauthorblockN{Zhe Jiang\IEEEauthorrefmark{2}, Minli Liao\IEEEauthorrefmark{5},Sam Ainsworth\IEEEauthorrefmark{6} Dean You\IEEEauthorrefmark{2}, Timothy Jones\IEEEauthorrefmark{5}}
\IEEEauthorblockA{\IEEEauthorrefmark{2}National Center of Technology Innovation for EDA, School of Integrated Circuits, South East University, People's Republic of China,\\
\IEEEauthorrefmark{5}University of Cambridge, United Kingdom,
\IEEEauthorrefmark{6}University of Edinburgh, United Kingdom}
\thanks{*The paper has been accepted by DAC'25. }
}

\maketitle

\begin{abstract}
Heterogeneous parallel error detection is an approach to achieving fault-tolerant processors, leveraging multiple power-efficient cores to re-execute software originally run on a high-performance core.
Yet, its complex components, gathering data cross-chip from many parts of the core, raise questions of how to build it into commodity cores without heavy design invasion and extensive re-engineering. 

We build the first full-RTL design, MEEK, into an open-source SoC, from microarchitecture and ISA to the OS and programming model.
We identify and solve bottlenecks and bugs overlooked in previous work, and demonstrate that MEEK offers microsecond-level detection capacity with affordable overheads.
By trading off architectural functionalities across codesigned hardware-software layers, MEEK features only light changes to a mature out-of-order superscalar core, simple coordinating software layers, and a few lines of operating-system code.
The Repo. of \name{}'s source code:
\textbf{\url{https://github.com/SEU-ACAL/reproduce-MEEK-DAC-25}}
\end{abstract}



\thispagestyle{plain}
\pagestyle{plain}

\section{Introduction}
\label{SC:Intro}

Hardware faults, both permanent and transient, can induce system anomalies and execution errors, which become more common with the increasing number of transistors and lower operating voltages in modern processors~\cite{hernandez2015timely,austin1999diva,romanescu2008core,reinhardt2000transient,snir2014addressing,yang2024efficient}.
To mitigate the errors caused by hardware faults, protection mechanisms exist, ranging from error codes to fault-tolerant architectures.
Detection is always key:
once an error is detected, the system can transition to a safe state, enabling corrective actions (\eg, system recovery or fault isolation).
As mandated by global safety standards from industry, \eg, ISO26262 for automotive~\cite{iso201126262} and DO-178C~\cite{rierson2017developing} for avionics, hardware faults must be addressed before escalating into hazards, \ie, within the Fault Tolerance Time Interval (FTTI), often measured in milliseconds~\cite{jiang2018bluevisor}.


Software mechanisms (\eg, multithreading~\cite{reinhardt2000transient,mukherjee2002detailed} and software scanner~\cite{bernardi2015development,batcher1999instruction}) typically incur significant performance degradation or offer limited fault coverage~\cite{michalak2005predicting}, making them insufficient for processors that require stringent reliability standards (\eg, ASIL-D in ISO26262~\cite{iso201126262}).
Hardware mechanisms often employ a dedicated core to execute a program copy, enabling the comparison of the core’s pins at each clock cycle (\eg, locksteps~\cite{iturbe2016triple,iturbe2019arm,stolt2008design}). 
By replaying everything on a separate, synchronized core and performing run-time verification at the signal level, full coverage and real-time guarantees are achieved.
Although dual- and triple-core locksteps have been successfully applied to microcontroller-sized processor cores in many life-critical application scenarios~\cite{RenesasRH850,Infineon_TriCore}, they have been shown impractical for Out-of-Order (OoO) superscalar cores due to prohibitive energy, area, and thermal costs~\cite{veeraraghavan2012doubleplay,gupta2008stagenet,gao2015survey}.

\parlabel{Heterogeneous parallel error detection.}
As a promising alternative, heterogeneous parallel error detection~\cite{ainsworth2018parallel,ainsworth2019paramedic,austin1999diva} leverages strong induction to divide a software program running on an OoO superscalar high-performance core (big core) into multiple discrete segments using Register Check Points (RCPs) and re-execute them on sets of smaller, power-efficient cores (little cores) for verification.

To replay memory and other non-repeatable operations, the addresses and data of relevant instructions (\eg, \texttt{load} and \texttt{store}) are extracted from the program stream at the commit stage within the big core, generating a partitioned, distributed log of load and store operations.
When the log of a segment is filled, or an instruction timeout is reached, a new RCP is triggered, and the corresponding little core begins verifying the segment between Start RCPs (SRCPs) and End RCPs (ERCPs).
By overlapping verification jobs, little cores collectively offer sufficient computational capacity to keep pace with the big core, ensuring full coverage with low overheads ~\cite{ainsworth2018parallel}.


\begin{figure}[t]
    \centering
    \includegraphics[width=0.955\columnwidth]{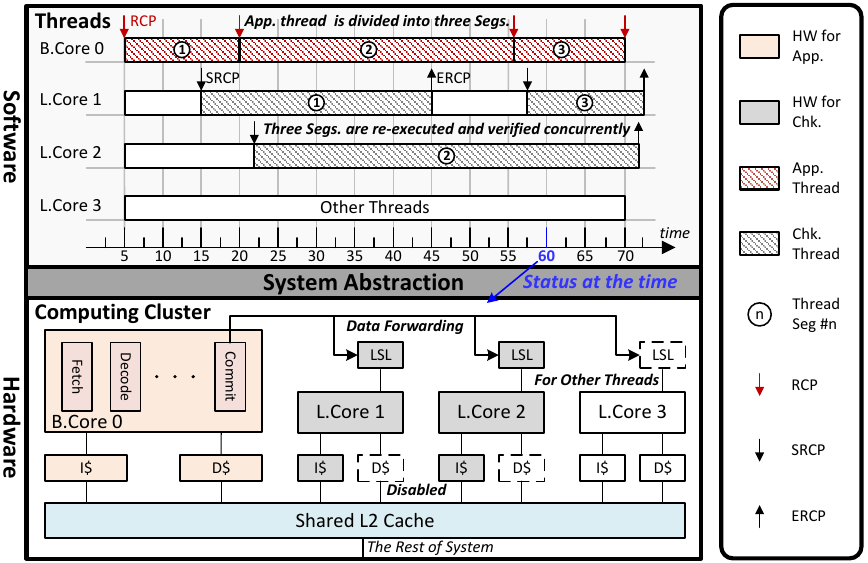}
    \caption{Re-constructed heterogeneous parallel error detection architecture \textit{(RCP: Register Checkpoint; S/ERCP: Start/End RCP LSL: Load-Store Log)}: an application thread on big core 0 is divided into three Segs. using RCPs, replayed and verified on little core 1 and 2.}
    \label{fig:Concepts}
\end{figure}

\parlabel{Challenges.}
Unlike conventional lockstep cores, which are typically implemented using identical processor cores with multiplexers and comparators~\cite{de2018lockstep}, heterogeneous error detection relies on intricate asynchronous interactions between the big core and a sea of little cores.
This involves continuously collecting and distributing loads, stores, and periodical register checkpoints from a big core and flexibly managing little cores.
This data needs to be preserved from execution time until commit, before transmission in order, and handling superscalar commit.
This also means that the collected data must be prioritized and routed to the corresponding little cores at high bandwidth, to prevent backpressure from stalling the big core. 

Existing work~\cite{ainsworth2018parallel,ainsworth2019paramedic} on the architecture has been studied through abstract simulation, modeled upon idealistic assumptions in microarchitecture and without an OS, imposing barriers for practical deployment. 
For instance, it is unclear how to collect such massive data from different locations of a real core without heavy microarchitectural changes, route it cross-chip while handling contention, and provide a configurable framework for little cores' management.


\parlabel{Contributions.}
We show it is possible to build heterogeneous error detection into real cores with minimal changes, by re-constructing the concept via hardware/software codesign, trading off functionalities across system layers.
This allows an application thread on a big core to be segmented using RCPs, replayed, and verified using any number of checker threads deployed on any little cores while still allowing other threads to be executed (Fig.~\ref{fig:Concepts}).
We build a  full-stack framework, \textbf{Make Each Error Count (MEEK)}, upon an open-source SoC \cite{asanovic2016rocket},  demonstrating the architecture can be realized with minimal design invasion within mature cores, and a few line changes in a full Linux.


\begin{figure*}[t]   
    \centering
    \includegraphics[width=1\textwidth]{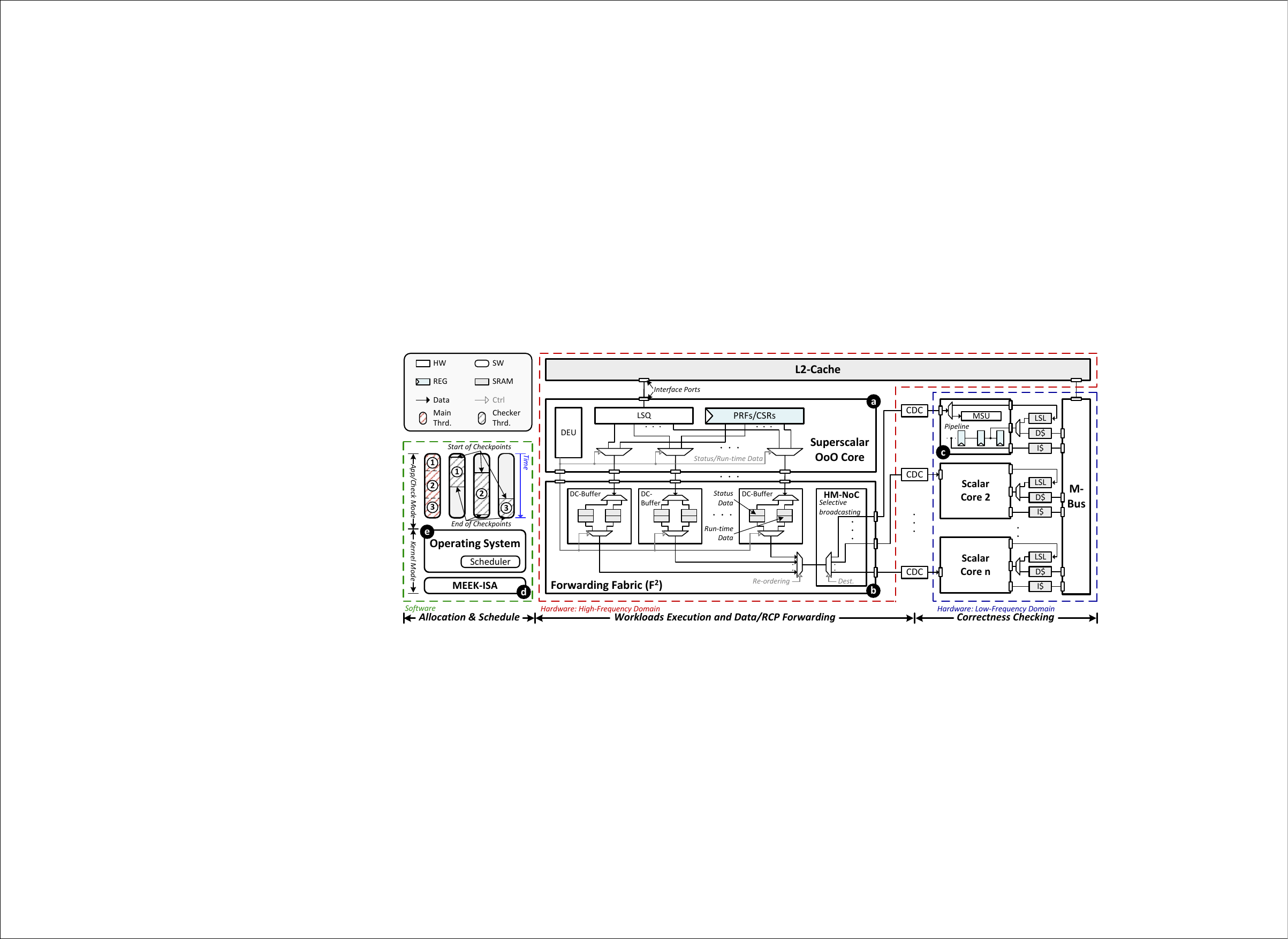}
    \caption{An overview of \name. \textit{(DEU: Data Extraction Unit; LSQ: Load-Store Queue; PRFs: Physical Register Files; CSRs: Control and Status Registers; HM-NoC: Half-duplex Multicast Network-on-Chip; MSU: Mode Switch Unit; LSL: Load-Store Log; M-Bus: Memory Bus)}. At hardware: \circled{a} a non-intrusive DEU is deployed in the big core, collecting data from various locations without disrupting the core's execution;
    \circled{b} a bespoke forwarding fabric F$^2$ is developed, prioritizing and distributing the data only to the relevant little cores;
    \circled{c} dual-mode little cores are designed for correctness checking or workload execution.
    At software: \circled{d} a customized ISA abstracts the control interface for \circled{e}, the scheduler in the OS, enabling flexible management of verification and workload execution on the little cores.
    }
    \label{fig:AnOverview}
\end{figure*}

\section{\name: A CPU/OS Codesigned Approach}
\label{sc:SA}
To implement MEEK at minimal complexity while achieving high performance, we had to make careful design-choice partitions between hardware and software (\cref{fig:AnOverview}). 
On one hand, we had little choice but to implement load- and store-logging, for replay of execution on little cores, in hardware: using software-based instrumentation~\cite{COMET} would have meant infeasibly high overheads. 
On the other hand, since little cores perform page-table walks asynchronously with the big core, the OS had to be aware of them in some form, and so we let the OS fully control their scheduling, avoiding implementing complex decisions that were not on the performance critical path in hardware, and allowing little cores to execute standard processes as well. Rather than a fully transparent interface, and to avoid expensive full error correction in hardware~\cite{ainsworth2019paramedic}, programs interact with the MEEK-ISA by being wrapped with coordinator functions inserted before main, which request checker resources from the OS, verify checking outputs, and call fault-handling code if needed. 

To do so, we slightly modified the big core's microarchitecture to insert a read-only observation channel at the commit stage (Fig.~\ref{fig:AnOverview} \circled{a}), collecting the big core's \emph{status data} (\ie, architectural, control and status register files) at each RCP and the \emph{run-time data} (addresses and data of memory and other non-repeatable operations) between RCPs.
We built a dedicated data fabric (Fig.~\ref{fig:AnOverview} \circled{b}), selectively broadcasting/routing the extracted data to the little core(s),
minimizing the backpressure from the data communications.
In little cores, the received data is buffered in a Load-Store Log (LSL), replacing the L1 cache during program replay, allowing the little core to reset its architectural state to a given SRCP, replay the exact instructions between the RCPs, and verify execution correctness at the ERCP, using the different types of data (Fig.~\ref{fig:AnOverview} \circled{c}).
Our ISA interface (Fig.~\ref{fig:AnOverview} \circled{d} and Tab.~\ref{table:ISA}) (re-)configures the little cores' checking characteristics, \ie, the operational mode  (application or check mode). 


\parlabel{Detection approach.}
Error checking is parallelized using checker threads (Fig.~\ref{fig:AnOverview} (b)): 
an application thread is segmented using RCPs,  taken when the targeted LSL is full, an instruction timeout is reached, or the kernel mode is trapped.
These segments are run in a second time by checker threads, assuming all previous segments are correct. 
After re-execution, the checker thread compares its architectural registers against the ones provided by the application thread at the same RCP. 
If the registers match, the segment is considered correct. 
If all segments pass check, the entire execution is deemed correct\footnote{ Memory operations, however, cannot be verified at ERCPs.  In the re-execution, addresses and data of memory operations are compared directly in the LSL. 
A similar design method is used for Control and Status Registers (CSRs) to verify non-repeatable instructions.}.

\begin{table}[t]
\caption{\name\ ISA \emph{(Priv 1/0: kernel/user modes)}.}
\label{table:ISA}
\centering
\resizebox{.95\columnwidth}{!}{%
\begin{tabular}{l|c|l}
\bottomrule
\hline
\textbf{Instruction} & \textbf{Priv} & \multicolumn{1}{c}{\textbf{Description}} \\ \hline
\texttt{\textbf{b.hook}} rs1, rs2      & 1             & Hook big core rs1 with little core rs2.   \\
\texttt{\textbf{b.check}} rs1         & 1             & Enable/Disable checking capacity.        \\
\hline
\texttt{\textbf{l.mode}} rs1, rs2      & 1             & Switch little core rs1's mode to rs2.    \\ 
\texttt{\textbf{l.record}} rs1         & 0             & Record arch. registers to address rs1.   \\
\texttt{\textbf{l.apply}} rs1          & 0             & Apply arch. registers from address rs1.  \\
\texttt{\textbf{l.jal}} rs1           & 0             & Jump to rs1 (PC of main thread).         \\
\texttt{\textbf{l.rslt} rd}               & 0             & Return the check results.                \\ \hline
\toprule
\end{tabular}
}
\end{table}

\section{The Microarchitecture}
\label{sc:Micro}

Even with the reduced hardware feature-set discussed in \cref{sc:SA}, \name{} required careful microarchitectural engineering to allow efficient, low-overhead execution, particularly around data transmission. We identify and ameliorate many key bottlenecks that were missing from the analyses and high-level simulations presented by previous work~\cite{ainsworth2018parallel,ainsworth2019paramedic,ainsworth2021paradox} due to a lack of RTL implementation. We also identify inefficiencies in terms of redundant data storage, where much of the information required by the forwarding paths is already buffered inside the core until commit time, meaning dedicated structures are not needed; instead we can forward data from the existing structures. Despite bottlenecks, we demonstrate that it is possible to build \name's microarchitecture from a mature heterogeneous SoC with very minor changes, avoiding heavy engineering efforts. 

We build \name{} into an open-source heterogeneous SoC (Rocket Chip~\cite{asanovic2016rocket}), with both high-performance and energy-efficient cores (BOOM and Rocket). 
We leverage Rocket Chip to demonstrate the applicability of our methodology to the other heterogeneous SoCs, \eg, ARM's big.LITTLE~\cite{big-little} and Intel's P- and E-class cores~\cite{doweck2017inside}.

\begin{figure}[t]
    \centering
    \includegraphics[width=1\columnwidth]{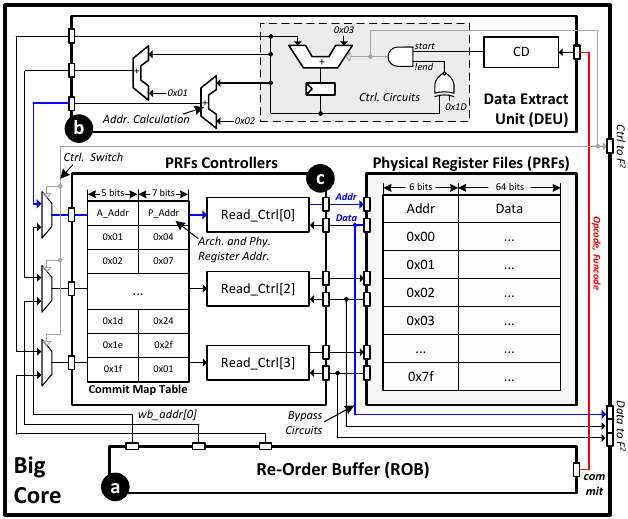}
    \caption{Big core microarchitecture, extracting status data 
    \textit{(red: ROB to DEU; blue: DEU to PRFs)}: 
    \circled{a} at commit time, opcode and function code are routed; \circled{b} DEU determines whether to extract status data; \circled{c} if so, a signal is routed and preempts the PRF controller for reading.}
    \label{fig:Main}
\end{figure}

\subsection{OoO Superscalar Core (The Big Core)}
\label{sc:BigCore}

Program replay requires the collection of both status data and run-time data from the big core.
The data is temporarily stored within the core during the program executions but distributed across various locations.
The status data is buffered in the Physical Register Files (PRFs) and Control and Status Registers (CSRs), while the run-time data is stored in the Load-Store Queue (LSQ)\footnote{By building MEEK, we noticed that Ainsworth et al.~\cite{ainsworth2018parallel}'s policy, of storing a dedicated buffer of load data, indexed by Re-Order Buffer (ROB) entry, was unnecessary. 
All the need was already buffered in LSQ. 
Yet, light changes were needed to preserve redundancy and thus fault tolerance, as while we assume data is protected via parity in the cache~\cite{schroeder2009dram}, and via full duplication by the point of reaching the LSL, there is a small window in the LSQ where it is otherwise protected by neither. 
We copy the cache's parity bits into the LSQ, double-checking them once the data is forwarded.
} and CSRs. All can be directly extracted but must occur in a narrow window, at commit time rather than execution time to reflect ordering, and before being overwritten by next instructions.  In light of this, we develop a \emph{non-intrusive} DEU to establish an observation channel (Fig.~\ref{fig:Main}), containing a Commit Detector (CD), control circuits and bypass circuits, injected at the PRFs, CSRs, and LDQ.
The CD monitors instruction commits from the ROB and selects the bypass circuits to extract data at RCPs or run-time data between RCPs.
This enables timely data extraction without bringing extra buffers or altering existing register paths.

\parlabel{PRF example.}
Fig.~\ref{fig:Main} reveals the microarchitecture using the PRFs (used to collect RCPs) as an example, where the DEU interfaces the ROB, PRFs, and F$^2$. 
At each instruction's commit, the opcode and function code are routed from the ROB (Fig.~\ref{fig:Main} \circled{a}) to the CD, allowing the CD to determine whether to extract data (Fig.~\ref{fig:Main} \circled{b}). 
If an RCP is reached, a signal is generated by the control circuits, preempting the PRF controller to read register files and forward them to the F$^2$ (Fig.~\ref{fig:Main} \circled{c}).
The PRF controllers are multiplexed between the ROB and the DEU, where the DEU has priority access when required, enabling immediate data reading and preventing data overwriting\footnote{A similar microarchitecture is implemented for the CSRs, allowing the data extraction from arbitrary CSR addresses. 
In contrast, because the top of the LSQ consistently holds data from the most recently retired instructions, when the CD decides to forward the run-time data, the bypass circuits directly transmit from the queue top, minimizing the design complexity.}.

\subsection{Forwarding Fabric (F$^2$)}
\label{sc:F2}
Previous simulations adopted a naive model for data forwarding, ignoring contention and critical paths~\cite{ainsworth2018parallel,ainsworth2019paramedic,hameed2010understanding,restuccia2020axi}.
Our attempt to implement such a mechanism revealed a stumbling block impeding system performance, even with the deployment of a full-featured AXI interconnect~\cite{jiang2022axi, jiang2022bluescale} (Sec.~\ref{sbsc:BottleneckAnalysis}).
This is caused by the huge amount of run-time data generated by the big core's parallel commits, in congestion with the frequent reaches of RCPs, which tends to occur in bursts, particularly when the core retires multiple \texttt{load} or \texttt{store} instructions at RCP boundaries, requiring many data transfers within a single cycle and high throughput.

We design F$^2$ with Dual-Channel Buffers (DC-Buffers) and a Half-duplex Multicast NoC (HM-NoC), storing and routing the extracted data  (Fig.~\ref{fig:AnOverview} \circled{b}).
A DC-Buffer is connected to each commit path, adding independent FIFOs for status and run-time data.
This ensures that all run-time data can be stored at the same cycle of commit, even when status data is generated simultaneously, avoiding data needing to be stored inside the core's own structures for longer than in the original core design.
HM-NoC uses a half-duplex (1-to-N) Manhattan grid. To achieve high enough throughput, this NoC allows the transmission of two packets per cycle, while preserving ordering\footnote{
Unlike previous work, where all data of a segment is buffered and forwarded collectively at an RCP, F$^2$ enables immediate data transmission and use upon collection time, allowing earlier re-execution by the little core.
Also, as the same status data might be required by two little cores (respectively used as the SRCP and ERCP), the data is selectively broadcast to the little cores when they are capable of data receiving, eliminating redundant transactions.}.

\begin{figure}[t]
    \centering
    \includegraphics[width=1\columnwidth]{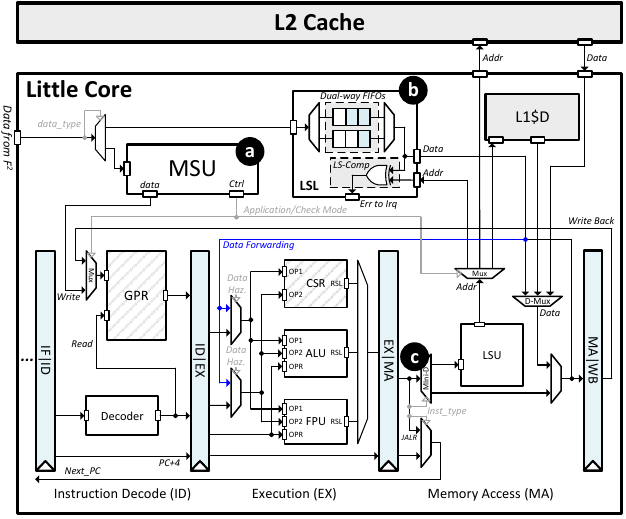}
    \caption{Little core microarchitecture, upgraded with the MSU and LSL:  the \circled{a} MSU servers as the control engineer, and the \circled{b} LSL buffers packets from F$^2$.
    While running the checker thread, the MSU manages its status, and memory data is fetched from the LSL.
    }
    \label{fig:Checker}
\end{figure}

\subsection{In-order Scalar Core (The Little Core)}
\label{sc:LittleCore}

To enable thread-level error detection and allow the co-existence of the checker and application threads, the little core's microarchitecture must support different operational modes and abstract an interface for software control.
Hence, we upgrade the little core's microarchitecture (Fig.~\ref{fig:Checker}) with a Mode Switch Unit (MSU, Fig.~\ref{fig:Checker} \circled{a}) and a Load-Store Log (LSL, Fig.~\ref{fig:Checker} \circled{b}).
The MSU serves as the control engine, and the LSL buffers received packets. 
While replicating the big core's states, the MSU records the little core's architectural register files and replaces them with the status data from the LSL.
The MSU also manages the little core's operational mode by inspecting the Thread ID (TID) of the running thread with that of the checker thread.
When the checker thread is scheduled, the MSU switches the operational mode, where the results of \texttt{load} and non-repeatable instructions are fetched from the LSL. 
Given that the little core accesses the LSL in an in-order manner, we implement the LSL bank using dual-way FIFOs, reducing complexity than conventional way-associative architectures.

\setlength{\textfloatsep}{-2pt}
\begin{algorithm}[t]
{\small
\SetAlgoLined
$\vartriangleright$ \text{\texttt{Scheduler}}\\
\SetKwFunction{FMain}{\normalfont Context\_Switch}
    \SetKwProg{Fn}{Function}{:}{}
    \Fn{\FMain{\normalfont task \textit{*current}, core \textit{core\_index}}}{
    {
        \textcolor{blue}{\textbf{\name}.b.check(DISABLE);}\\
        \textbf{Kernel}.Intr(DISABLE);\\
        task *\textit{next} = NULL;\\
        
        \emph{/* switching current task to the next task */}\\
        \textbf{Kernel}.Context.save(\emph{current});\\
        \emph{next} = \textbf{Kernel}.Find\_next();\\
        \eIf {\normalfont (\emph{next$\rightarrow$new\_release})}{
            \For{$i = 0$ \KwTo \normalfont(size\_of(\emph{next$\rightarrow$checker\_index}) $- 1$)}{
                \emph{/* hook little cores to the big core */}\\
                \textcolor{blue}{\textbf{\name}.b.hook(\emph{core\_index}, \emph{next$\rightarrow$checker\_index}[i]);}\\
            }
           \textbf{Kernel}.Context.init(\emph{next});\\
        }
        {
            \textbf{Kernel}.Context.restore(\emph{next});\\
        }
        \emph{current} = \emph{next};\\
        \textbf{Kernel}.Intr(ENABLE);\\
        \textcolor{blue}{\textbf{\name}.b.check(ENABLE);}\\
        \textbf{Kernel}.Context.jalr(\emph{current$\rightarrow$pc});
    }
}
\textbf{End Function}
}
\caption{Big core's context switch \emph{(Blue: added code)}.}
\label{al:SW-B}
\end{algorithm}

\parlabel{Pipeline integrations.}
Fig.~\ref{fig:Checker} shows the integration of LSL and MSU into a 5-stage pipelined little core.
The LSL is added into the Memory Access (MA) stage (Fig.~\ref{fig:Checker} \circled{b}) by deploying a multiplexer and connecting it to the address port of the Load-Store Unit (LSU).
The multiplexer selectively routes memory accesses to the LSL based on the operational mode (returned by the MSU) and combines the virtual index and physical tag (returned by the TLB) into the address port of the LSL.
Also, a demultiplexer is integrated into the MA stage to direct the read data back to the next stages. 
A pair of multiplexer and demultiplexer are deployed into the Instruction Decode (ID) stage to allow the recording and updating of the architectural registers.
Lastly, we deploy a Mini-Decoder (Mini-D, Fig.~\ref{fig:Checker} \circled{c}) at the MA stage to differentiate conventional RISC-V and MEEK-ISA.

\parlabel{Performance-gap mitigation.}
In previous work~\cite{ainsworth2018parallel,ainsworth2019paramedic,ainsworth2021paradox}, a lack of microarchitectural detail meant bridging the performance gap between the big and little cores involved merely scaling the core count of the little cores. Unfortunately, when we build MEEK using real RTL, the Rocket little cores are each a significant fraction of the size of the big core, and thus using twelve of them as in the original deisgn~\cite{ainsworth2018parallel} would be infeasible (\cref{sc:area}). 
In particular, we discovered that there was a wide variety in the little cores' ability to keep up with the big core depending on instruction distribution~\cite{li2009mcpat}. 
With this, we realized that the best way to minimize core overhead while achieving good performance was not to use the smallest Rocket cores available, but rather to \textit{balance} their bottlenecks with respect to BOOM. 
To reduce the performance gap, we optimize little cores by increasing the size of bottlenecked components, \eg, increasing divider unrolling and extending FPU pipeline.

\section{The ISA, Operating System, and Programming Model}
\label{sc:SW}

We detail ISA support needed to add to augment programs with fault-tolerance support.
With just a few lines-of-code changes to the kernel, it can schedule and reserve resources for checker threads while allowing standard scheduling and context switching of other threads.

\subsection{ISA Support}
\label{sbsc:ISA}

The new ISA is classified into two categories (Tab.~\ref{table:ISA}) for the big core (\texttt{b.x()}) and the little core (\texttt{l.x()}).
We use \texttt{b.hook()} to set the association between the big and little cores, followed by \texttt{b.check()} to enable/disable the checking capacity via switching on/off the DEU.
For little core, \texttt{l.mode()} sets its operational mode, and a pair \texttt{l.record()} and \texttt{l.apply()} record and apply the architectural registers from a given source. 
To re-direct the PC to an application thread, we develop a \texttt{l.jal()}, modified from the vanilla jump instruction with an alteration on the target.
By treating a checkpoint-end as branch-like, the pipeline handles control hazards from the PC-change, without further changes.  
Lastly, \texttt{l.rslt()} indicates whether an RCP mismatch is detected.
As \texttt{b.hook()} and \texttt{b.check()} can lead to contention in use of the little cores, and the \texttt{l.mode()} can cause erroneous execution from unintended memory accesses, they are privileged instructions, executed via OS syscall.

\subsection{Checker Thread and Its Programming Model}

The checker thread is initialized with the application thread by augmenting the application thread's main function using constructor and destructor functions~\cite{IBM} (Al.~\ref{al:SW-B}: line 14).
Since the checker thread relies on the data in LSL to replay memory operations and the LSL is designed using FIFOs, context switches of the log are undesirable.
Hence, during the scheduling time, LSL is reserved for a single checker thread (Al.~\ref{al:SW-B}: line 12), even if multiple threads can be scheduled on the core.
Once LSL is reserved, only data relevant to the associated checker thread is forwarded until re-execution is complete.
Likewise, a checker thread pinned to a specific application thread cannot be migrated before the re-execution completes 
Since each checkpoint is finite in size (5000-instruction maximum), and since ownership returns to the OS after the end of each checkpoint for reallocation, this does not cause resource starvation.

\begin{algorithm}[t]
{\small
\SetAlgoLined
$\vartriangleright$ \text{\texttt{Scheduler}}\\
\SetKwFunction{FMain}{\normalfont Context\_Switch}
    \SetKwProg{Fn}{Function}{:}{}
    \Fn{\FMain{\normalfont task \textit{*current}, core \textit{core\_index}}}{
    {
        \textcolor{blue}{\textbf{\name}.l.mode(MODE\_APPLICATION);}\\
        \emph{/* switching current task to the next task */}\\
        ... \\

        \If {\normalfont (\emph{next$\rightarrow$checker\_thread})}{
            \textcolor{blue}{\textbf{\name}.l.mode(MODE\_CHECK);}
        }
        \textbf{Kernel}.Context.jalr(\emph{current$\rightarrow$pc});
    }
}
\textbf{End Function}\\
}
\small
$\vartriangleright$ \text{\texttt{Checker Thread, Newly Developed}}\\
\SetKwFunction{FMain}{\normalfont Checker\_Thread}
    \SetKwProg{Fn}{Function}{:}{}
    \Fn{\FMain{}}{
    {
        \emph{/* initializing checker thread using P-Thread */}\\
        ... \\
        \normalfont \textbf{\name}.l.record(\emph{sp});  \emph{// after checking, returns here}\\
        \If {\normalfont (!\textbf{\name}.l.rslt())}{
            \textbf{\name}.ReportErr();
        }

        \textbf{while} {\normalfont (\textbf{\name}.NewSRCP()$\rightarrow$invalid);} \\
        \textbf{\name}.l.apply(\emph{LSL});\\
        \textbf{\name}.l.jal(\textbf{\name}.NewSRCP()$\rightarrow$pc);
    }
    
}
\textbf{End Function}
\caption{Little core's context switch and checker thread.}
\label{al:SW-L}
\end{algorithm}

\parlabel{Programming model.}
We develop the checker thread based on the new ISA, ensuring minimal coding efforts: 
initially, an \texttt{l.record()} is employed to record the current architectural register state (Al.~\ref{al:SW-L}: line 15), allowing the core to return after verification. 
Then, a busy loop is created to await the arrival of status data in the LSL (Al.~\ref{al:SW-L}: line 19).
Upon receiving it, \texttt{l.apply()} is invoked to modify the core's architectural state in accordance with the application segments (Al.~\ref{al:SW-L}: line 20), with a \texttt{l.jalr()} being invoked to redirect PC to the replicated target (Al.~\ref{al:SW-L}: line 21).
Lastly, a \texttt{l.rslt()} is used to return the verification status. 
If an error is detected, an interrupt is triggered to notify the OS for corrective actions (Al.~\ref{al:SW-L}: line 16 - 18).

\subsection{OS Kernel and Its Verification}
\label{sbsc:OSKernel}

With the new ISA, kernel modifications can be constrained to the context switch function within the scheduler, allowing the configuration of checking capabilities and the management of checker threads~\cite{aas2005understanding,love2010linux}.
In accordance with the distinct roles of the big and little cores, the context switches for application and checker threads are modified differently.
While entering the context switch from an application thread, \texttt{b.check(DISABLE)} is invoked to deactivate the checking functionality (Al.~\ref{al:SW-B}: line 3), and upon leaving the context switch, \texttt{b.check(ENABLE)} is called to reactivate it (Al.~\ref{al:SW-B}: line 20).  
Furthermore, if a newly released thread is scheduled for execution, the \texttt{b.hook(core\_index, next$\rightarrow$checker\_index[x])} is used to associate it with the little cores (Al.~\ref{al:SW-B}: lines 10 - 13). 
For the context switch of application threads, the only required modification is using the \texttt{l.mode()} to switch the little core's operational mode (Al.~\ref{al:SW-B}: lines 3 and 7), leaving the remainder unchanged.

\begin{figure}[h]
    \centering
    \includegraphics[width=1\columnwidth]{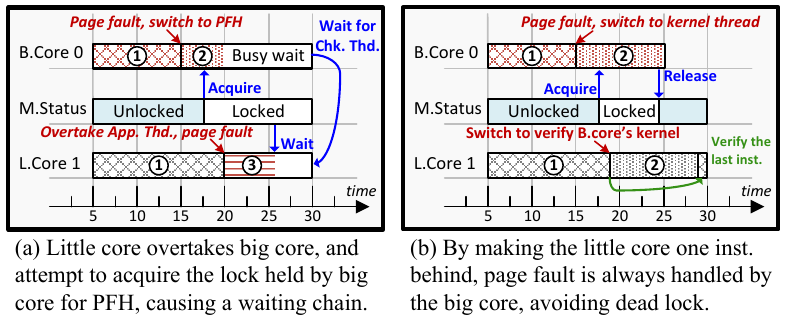}
    \caption{A deadlock occurs when the little core tries to acquire the lock held by the big core. By building synchronization between cores and I/Os, preventing the checker ever needing to claim locks, the deadlock is resolved \emph{(M.Status: Memory Status; PFH: Page Fault Handler)}. }
    \label{fig:Deadlock}
\end{figure}

\parlabel{Kernel verification and deadlock resolving.}
Since \name\ enables error checking at the thread level, the OS kernel can be treated as a specialized application thread and verified like the other application threads. 
Yet, during the development, we observed a \textbf{\textit{deadlock}} missed in previous literature~\cite{ainsworth2018parallel,ainsworth2019paramedic,ainsworth2021paradox} due to a lack of evaluation with OS. 
Because a checker thread can block a main thread until the little core has finished, due to SRAM logs being finite, this behavior acts as a lock that the little core holds and the big core needs.
If there are scenarios where the opposite occurs (the big core holds locks, e.g., software mutexes, required by the little core), then deadlock results. 
This is impossible during standard re-execution, as the checker thread itself does not access memory and thus cannot take locks; only replaying the memory previously read when the main thread took out locks.
However, there are scenarios where the checker thread inadvertently requires real memory reads in order to continue: specifically, if an instruction fault on the little core (e.g., because it has overtaken the big core, or the instruction was paged out before the little core reaches it)~\cite{bovet2005understanding}, the page-fault handling may require a lock held by the main thread, causing deadlock (Fig.~\ref{fig:Deadlock} (a)).

This deadlock can be fixed by making it impossible for a checker thread to fault on instructions. Ensuring the checker thread is at least one instruction behind the main thread means the latter will reach faults first. Synchronizing on I/O makes it impossible to write out a page potentially used by an unfinished checker thread (Fig.~\ref{fig:Deadlock} (b))\footnote{More generally, care needs to be taken if kernel operations are inside the sphere of replication: similar effects could happen were the scheduler blocked via a lock held by a waiting main thread, in turn blocking checkers.}.

\section{Evaluation}
\label{sc:Evaluation}
We built \name\ upon an open-source heterogeneous SoC (Rocket Chip~\cite{asanovic2016rocket}), featuring both high-performance (BOOM~\cite{zhao2020sonicboom}) and energy-efficient (Rocket~\cite{asanovic2016rocket}) cores.
The BOOM was augmented with DEU and F$^2$ as the big core, and the Rocket was upgraded with LSL and MSU to support re-execution as the little core.
We implemented the microarchitecture with Chisel (v3.4) and synthesized the RTL using Vivado toolchains (v2021.2). 
The generated netlist was deployed on AMD UltraScale+ FPGAs using FireSim~\cite{karandikar2018firesim}, emulating the setup in Tab.~\ref{table:Setup}.
We booted Linux kernel v5.7, executing full SPECint 06~\cite{henning2006spec} and Parsec V3~\cite{bienia2008parsec} with simmedium dataset.


\subsection{Performance Overhead}
\label{sbsc:PerformanceOverhead}
\begin{figure*}[t]
    \centering
    \includegraphics[width=1\linewidth]{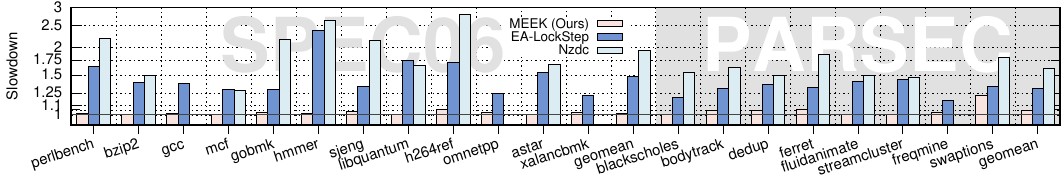}
    \caption{Performance results for \name\ (4 little cores), Equivalent-Area Lockstep (EA-LockStep) and Nzdc~\cite{didehban2016nzdc}, running SPEC06 and Parsec.}
    \label{fig:Perf}
\end{figure*}

Fig.~\ref{fig:Perf} shows the slowdown experienced by the big core when running SPECint and Parsec in \name, compared to the software-based counterparts implemented in LLVM (Nzdc~\cite{didehban2016nzdc})\footnote{For Nzdc, compilation fails in gcc, omnetpp, xalancbmk, and freqmine.} and  LockStep with Equivalent Area (EA-LockStep). 
Nzdc was chosen for comparison as it is the only open-source software mechanism available, while lockstep is the most widely used hardware mechanism. 
However, simply duplicating the core in lockstep would consume double the area of the big core while achieving the same performance as a vanilla core, leading to uninteresting comparisons. 
Therefore, we scaled down the big core's configurations, through linear interpolation on each configurable BOOM component, to create a comparator with both cores combined matching the area overhead of MEEK.
In all cases, \name\ is configured with four little cores.

\begin{table}[]
\centering
\resizebox{.90\columnwidth}{!}{%
\begin{tabular}{ll}
\multicolumn{2}{c}{\textit{Big Core}}                                                                                                             \\ \hline
Core         & \begin{tabular}[c]{@{}l@{}}4-Width, OoO superscalar SonicBoom, @3.2GHz\\ 128-Entry ROB,96-entry IQ, 32-entry LDQ/STQ,\end{tabular} \\
Pipeline     & \begin{tabular}[c]{@{}l@{}}128 Int/FP Phy Registers, 2 Int ALUs, 1 FP/Mult/Div\\ ALU, 2 MEM, 1 Jump Unit, 1 CSR Unit\end{tabular}  \\
Branch Pred. & \begin{tabular}[c]{@{}l@{}}TAGE algorithm, 256-entry BTB, 32-entry RAS,\\ 6 TAGE table with 2 - 64 bits history\end{tabular}       \\
\multicolumn{2}{c}{\textit{Memory Hierarchy}}                                                                                                     \\ \hline
L1 ICache    & 32 KB, 4-way, 8 MSHRs                                                                                                              \\
L1 DCache    & 32 KB, 4-way, 8 MSHRs                                                                                                              \\
L2 Cache     & 512 KB, 8-way, 12 MSHRs                                                                                                            \\
LLC          & 4 MB, 8-way, 8 MSHRs                                                                                                               \\
Memory       & 16 GB DDR3 @1066Mhz, max 32 requests                                                                                               \\
\multicolumn{2}{c}{\textit{Little Cores}}                                                                                                         \\ \hline
Cores        & \begin{tabular}[c]{@{}l@{}}4 $\times$ In-order Rocket, 5-stage pipeline, @1.6GHz, \\ 8-Unroll DIV, 3-stage FPU\end{tabular}                   \\
LSL          & 4 KB, 5000 instruction time-out                                                                                                    \\
L1 Cache     & 4 KB, 2-way for both I- and D-Cache                                                                                                \\ \hline
\end{tabular}}
\caption{Hardware configurations evaluated. }
\label{table:Setup}
\end{table}

Using four little cores is sufficient to execute SPEC with overheads of 1.4\% and Parsec at 4.4\% (geomean).
For all workloads, except swaptions, the observed slowdown is below 5\%. 
Swaptions, however, suffers the highest slowdown of 22\%, due to the frequent divisions, where the Rocket core's divider is significantly less performant than the BOOM core's.
For the comparators,  Nzdc introduces geomean overheads of 60.2\% on Parsec and 94.2\% on SPEC, reflecting limitations associated with its software implementation. 
The hardware counterpart, EA-LockStep, incurred geomean overheads of 31.2\% for Parsec and 48.7\% for SPEC, approximately 6.1x and 33.7x higher than \name, evidencing the performance-area benefits of \name.

\subsection{Detection Latency}
\label{sbsc:DetectionLatency}

\begin{figure}[t]
    \centering
    \includegraphics[width=1\linewidth]{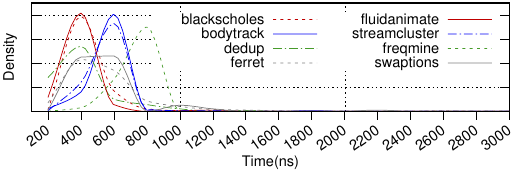}
    \caption{Detection latency while using 4 little cores (unit: $ns$).}
    \label{fig:density}
\end{figure}

To examine the detection latency, we inject errors in the forwarded data from the F$^2$ connected to the big core, \eg, data and address of memory operations and architectural register data, simulating the hardware faults without disrupting the big core's normal execution.
For each workload, 5,000 - 10,000 faults are randomly generated, and the density of detection latencies is shown in Fig.~\ref{fig:density}.

Each distribution features a long, but very thin, tail extending to the far right: the average detection latency is below 1 $\mu$s, while the worst-case latencies are 5 to 10 times higher, reaching up to 2.7 $\mu$s (in the ferret). 
Despite the randomness of the fault injection, the extremely high number of sample points ($\textgreater$ 100, 000, in total) suggests empirically that 3 $\mu$s is sufficient to cover over 99.9\% of hardware faults, which is multiple orders of magnitude lower than the millisecond-level FTTI requirement for ASIL-D compliance.

\subsection{Scalability}
\label{sbsc:Scalability}

\begin{figure}[t]
    \centering
    \includegraphics[width=1\linewidth]{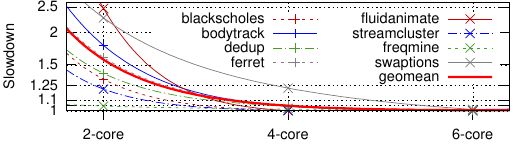}
    \caption{Slowdown when using varying numbers of little cores.}
    \label{fig:scale}
\end{figure}

Fig.~\ref{fig:scale} presents the slowdown observed with varying numbers of little cores. Using two little cores to verify the Parsec execution results in a 54.9\% slowdown (geomean), which remains significantly lower than that of Nzdc. With four little cores, the geomean overhead is reduced to 4.4\%, with only the swaptions workload experiencing more than 5\% overhead (as explained above). 
Scaling the system to six cores further decreases the geomean overhead to 0.3\%, with all workloads having less than 1\% overhead. 
The trend of slowdown exhibits a superlinear decline as the number of little cores increases, which suggests that adding little cores can effectively mitigate performance overhead even for more complex workloads in the future.

\begin{figure}[t]
    \centering
    \includegraphics[width=1\linewidth]{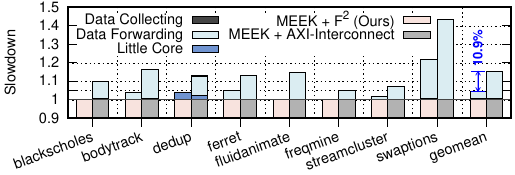}
    \caption{Backpressure decomposition (with 4 little cores).}
    \label{fig:bottleneck}
\end{figure}

\begin{figure}[t]
    \centering
    \includegraphics[width=1\linewidth]{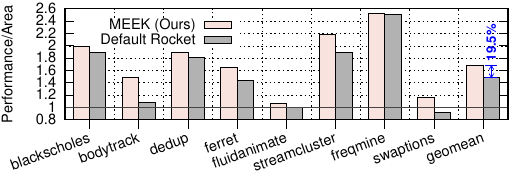}
    \caption{Performance-area analysis of the little core.}
    \label{fig:littleP}
\end{figure}

\subsection{Microarchitectural Bottleneck Analysis}
\label{sbsc:BottleneckAnalysis}
By detailed implementation, we identified the performance bottlenecks related to data collection, forwarding, and thread re-execution in the heterogeneous error detection process (Sec.~\ref{sc:Micro}).
Here, we provide an in-depth analysis of these bottlenecks and examine the effectiveness of our proposed microarchitecture.

\parlabel{Backpressure decomposition.}
Fig.\ref{fig:bottleneck} illustrates the overhead breakdown in \name\ using 4 little cores while running Parsec, employing a full-featured AXI-Interconnect~\cite{Xilinx_AXI} and our F$^2$.
As shown in Fig.~\ref{fig:bottleneck}, the AXI-Interconnect introduces a significant 16.7\% overhead (geomean), evidencing that the 128-bit narrow bus, only handling one packet per cycle, is the primary bottleneck of the entire system, particularly when the little cores  operating at a lower clock frequency. 
This also motivates the development of a dedicated data forwarding fabric with a wider data path and improved packets' scheduling to mitigate these limitations (Sec.~\ref{sc:F2}).
By replacing the AXI-Interconnect with a 256-bit fabric that supports multiple-packet transmissions per cycle while preserving ordering using FSMs, we effectively reduce the total overhead of data collection and forwarding to less than 5\%. 
Our F$^2$ design alleviates this bottleneck by enabling parallel data transmission and improved clock-domain crossing efficiency, thereby minimizing contention.
With F$^2$, \name\ shifts from being a system limited by data forwarding to one that is computation-bound, with performance depending on the combined compute power of its checker threads.

\parlabel{Little core's performance-area efficiency.}
As discussed in Sec.\ref{sc:LittleCore}, we analyzed to understand performance gaps between the little and big cores and increased the size of bottlenecked components in the little cores.
The evaluation shows that four optimized little cores can match the performance of six default little cores for the verification job. 
Given that our configurations increase the area of the little cores, we assess them using a normalized performance/area metric for fair comparison (Fig.\ref{fig:littleP}), revealing a 15.2\% improvement (geomean) and demonstrating the performance-area efficiency of our design choice.

\subsection{Hardware Overhead}
\label{sc:area}
We synthesized a physical implementation of \name\ with four little cores using TSMC 28nm PDKs~\cite{TSMC_28}.
The RTL is synthesized using Design Compiler (v2022.12), and the netlist is placed and routed via IC Compiler 2 (v2022.12), see Tab.~\ref{tab:HW_Overhead}.

The BOOM's area is 2.811 mm$^2$, and each Rocket is 0.092 mm$^2$, excluding L1 D\$ (since it is not required for re-execution~\cite{ainsworth2018parallel,ainsworth2019paramedic}).
The total area for BOOM's data collecting and forwarding takes 0.122 mm$^2$, (4.3\% of the BOOM): the DEU occupies 0.071 mm$^2$ and the F$^2$ consumes 0.051 mm$^2$.
In summary, building \name\ with four Rocket upon a BOOM needs 0.726 mm$^2$ in total, \ie, 25.8\% overhead.

\subsection{Gap Analysis}
\label{sc:gap}
We observe a discrepancy between the hardware overhead measured in MEEK and the estimation in previous high-level simulations~\cite{ainsworth2018parallel}.
While both work claims around 25\% extra area is enough to enable heterogeneous parallel error detection for a Cortex-A57-level core\footnote{IPCs of BOOM and A57 are similar, 0.95 and 0.97, measured using Parsec.}, this is concluded based on different configurations: twelve Rocket cores in~\cite{ainsworth2018parallel} versus four Rockets at double the frequency here.

Upon further analysis (Tab.~\ref{tab:HW_Overhead}), several key factors were identified.
First, BOOM is only 72.1\% the size of an A57, likely due to differences in ISA complexity.
This means that it is misleading to use Rocket to match the performance of an A57 in~\cite{ainsworth2018parallel}, as a
 little ARM core with similar performance would be larger than the RISC-V core.
Second, our implementation required 17.9\% more area per core than the synthesis results used in~\cite{ainsworth2018parallel}, where their L1 instruction cache configuration was also insufficient for large benchmarks like SPEC (also noted in~\cite{ainsworth2019paramedic}).
Lastly, wrapper logic was previously ignored. 

\begin{table}[!h]
\centering
\resizebox{0.99\linewidth}{!}{%
\begin{tabular}{ccccccc}
\bottomrule
\hline
\multicolumn{2}{c}{}                                                                                        & \multicolumn{1}{c}{\textbf{Big}}                 & \textbf{Little}                                    & \textbf{}                                              & \multicolumn{1}{c}{\textbf{Big}}                & {\color[HTML]{333333} \textbf{Little}} \\ \hline
\toprule

\multicolumn{1}{c|}{Core}                            & \multicolumn{1}{c|}{}                                & \multicolumn{1}{c|}{BOOM}                         & \multicolumn{1}{c|}{Rocket}                        & \multicolumn{1}{c|}{}                                  & \multicolumn{1}{c|}{Cortex-A57}                  & Rocket                                 \\
\multicolumn{1}{c|}{Number}                            & \multicolumn{1}{c|}{}                                & \multicolumn{1}{c|}{1}                            & \multicolumn{1}{c|}{\tikz[baseline=(char.base)]{\node[fill=gray!50, text opacity=1, fill opacity=0.5, rounded corners=3pt] (char) {4};}}     & \multicolumn{1}{c|}{}                                  & \multicolumn{1}{c|}{1}                           & \tikz[baseline=(char.base)]{\node[fill=gray!50, text opacity=1, fill opacity=0.5, rounded corners=3pt] (char) {12};}             \\
\multicolumn{1}{c|}{Freq. (GHz)}                            & \multicolumn{1}{c|}{}                                & \multicolumn{1}{c|}{3.2}                            & \multicolumn{1}{c|}{\tikz[baseline=(char.base)]{\node[fill=gray!50, text opacity=1, fill opacity=0.5, rounded corners=3pt] (char) {2};}}     & \multicolumn{1}{c|}{}                                  & \multicolumn{1}{c|}{3.2}                           & \tikz[baseline=(char.base)]{\node[fill=gray!50, text opacity=1, fill opacity=0.5, rounded corners=3pt] (char) {1};}             \\
\multicolumn{1}{c|}{Tech. (nm)}                      & \multicolumn{1}{c|}{}                                & \multicolumn{1}{c|}{28}                           & \multicolumn{1}{c|}{28}                            & \multicolumn{1}{c|}{}                                  & \multicolumn{1}{c|}{20}                          & 40                                     \\
\multicolumn{1}{c|}{Area (mm$^2$)}    & \multicolumn{1}{c|}{}                                & \multicolumn{1}{c|}{2.811}                         & \multicolumn{1}{c|}{0.092}                          & \multicolumn{1}{c|}{}                                  & \multicolumn{1}{c|}{2.050}                        & 0.160                                   \\
\multicolumn{1}{c|}{@28nm}                    & \multicolumn{1}{c|}{}                                & \multicolumn{1}{c|}{\tikz[baseline=(char.base)]{\node[fill=gray!50, text opacity=1, fill opacity=0.5, rounded corners=3pt] (char) {2.811};}} & \multicolumn{1}{c|}{0.092}                         & \multicolumn{1}{c|}{}                                  & \multicolumn{1}{c|}{\tikz[baseline=(char.base)]{\node[fill=gray!50, text opacity=1, fill opacity=0.5, rounded corners=3pt] (char) {3.905};}} & 0.078                                  \\
\multicolumn{1}{c|}{Wrapper (mm$^2$)} & \multicolumn{1}{c|}{}                                & \multicolumn{1}{c|}{\tikz[baseline=(char.base)]{\node[fill=gray!50, text opacity=1, fill opacity=0.5, rounded corners=3pt] (char) {0.122};}} & \multicolumn{1}{c|}{\tikz[baseline=(char.base)]{\node[fill=gray!50, text opacity=1, fill opacity=0.5, rounded corners=3pt] (char) {0.059};}} & \multicolumn{1}{c|}{}                                  & \multicolumn{1}{c|}{\tikz[baseline=(char.base)]{\node[fill=gray!50, text opacity=1, fill opacity=0.5, rounded corners=3pt] (char) {x};}}   & \tikz[baseline=(char.base)]{\node[fill=gray!50, text opacity=1, fill opacity=0.5, rounded corners=3pt] (char) {x};}              \\ \cline{1-1} \cline{3-4} \cline{6-7} 

\multicolumn{1}{c|}{\begin{tabular}[c]{@{}c@{}}\textbf{Overhead} \end{tabular}} & \multicolumn{1}{c|}{\multirow{-8}{*}{\rotatebox[origin=c]{270}{\textbf{Ours}}}} & \multicolumn{2}{c|}{\textbf{25.8\%}}                                                                            & \multicolumn{1}{c|}{\multirow{-8}{*}{\rotatebox[origin=c]{270}{\textbf{DSN'18~\cite{ainsworth2018parallel}}}}} & \multicolumn{2}{c}{\textbf{24\%}}                                                                  \\ \hline
    \toprule

\end{tabular}}
\caption{Hardware overhead in MEEK and DSN'18~\cite{ainsworth2018parallel}, excluding L1 D\$ in little cores (Grey shading indicates a key discrepancy). }
\label{tab:HW_Overhead}
\end{table}

Although the discrepancies in area estimation for individual modules may seem minor, the cumulative effect led to a significant overall mismatch between the simulated and real overhead, meaning we had the budget for only $\frac{1}{3}$ the little cores and so had to carefully optimize each one's performance (\cref{sbsc:BottleneckAnalysis}) rather than scaling the number of cores as was the perspective of the original papers.

\section{Conclusion}
\label{sc:Conclusion}
We have presented \name, a systematic implementation of heterogeneous parallel error detection, from the microarchitecture and ISA to the OS and programming model. 
While the typical architectural research employs abstract modeling (\eg, GEM5~\cite{binkert2011gem5}) for rapid prototyping, which is significantly faster than  real implementations, this work shows the practical challenges of implementing even a well-modeled architecture, identifying bugs, performance bottlenecks, and inaccurate estimations that are invisible in high-level simulations, as well as opportunities for heavy simplification via codesign. We believe there is much new insight to be gained at this level of analysis for the movement of complex research concepts into full production.




\section{Acknowledgement}
We appreciate the anonymous reviewers for their helpful feedback. 
This work is supported by the National Key Research and Development Program (Grant No. 2024YFB4405600), the National Natural Science Foundation of China (Grant No. 62472086, 62204036) and the Basic Research Program of Jiangsu (Grants No. BK20243042).


\bibliographystyle{IEEEtran}
\bibliography{refs}

\end{document}